\documentclass[twocolumn,aps,prb,floats,superscriptaddress,showpacs]{revtex4}
\usepackage{amsmath}
\usepackage{epsfig}
\usepackage{graphicx}
\usepackage{dcolumn}
\usepackage{bm}
\usepackage{color}

\def\gapp{\lower.35em\hbox{$\stackrel{\textstyle>}{\sim}$}}
\def\lapp{\lower.35em\hbox{$\stackrel{\textstyle<}{\sim}$}}

\begin{document}

\title{{Magnetically tunable multi-band near-field radiative heat transfer
between two graphene sheets}}
\author{Lixin Ge}
\email{lixinge@hotmail.com}
\affiliation{School of Physics and Electronic Engineering, Xinyang Normal University,
Xinyang 464000, China}
\affiliation{Division of Computer, Electrical and Mathematical Sciences and Engineering,
King Abdullah University of Science and Technology (KAUST), Thuwal
23955-6900, Saudi Arabia}
\author{Ke Gong}
\affiliation{School of Physics and Electronic Engineering, Xinyang Normal University,
Xinyang 464000, China}
\author{Yuping Cang}
\affiliation{School of Physics and Electronic Engineering, Xinyang Normal University,
Xinyang 464000, China}
\author{Yongsong Luo}
\affiliation{School of Physics and Electronic Engineering, Xinyang Normal University,
Xinyang 464000, China}
\author{Xi Shi}
\affiliation{Department of physics, Shanghai Normal University, Shanghai, 200234, China}
\author{ Ying Wu}
\email{ying.wu@kaust.edu.sa}
\affiliation{Division of Computer, Electrical and Mathematical Sciences and Engineering,
King Abdullah University of Science and Technology (KAUST), Thuwal
23955-6900, Saudi Arabia}
\date{\today}

\begin{abstract}
Near-field radiative heat transfer (NFRHT) is strongly related with many
applications such as near-field imaging, thermos-photovoltaics and thermal
circuit devices. The active control of NFRHT is of great interest since
it provides a degree of tunability by external means. In this work, a
magnetically tunable multi-band NFRHT is revealed in a system of two
suspended graphene sheets at room temperature. It is found that the
single-band spectra for B=0 split into multi-band spectra under an
external magnetic field. Dual-band spectra can be realized for a modest
magnetic field (e.g., B=4 T). One band is determined by intra-band
transitions in the classical regime, which undergoes a blue shift as the
chemical potential increases. Meanwhile, the other band is contributed by
inter-Landau-level transitions in the quantum regime, which is robust
against the change of chemical potentials. For a strong magnetic field
(e.g., B=15 T), there is an additional band with the resonant peak appearing
at near-zero frequency (microwave regime), stemming from the magneto-plasmon
zero modes. The great enhancement of NFRHT at such low frequency has not
been found in any previous systems yet. This work may pave a way for
multi-band thermal information transfer based on atomically thin graphene
sheets.
\end{abstract}

\maketitle

%\pacs{78.67.Pt, 73.20.Mf, 42.25.-p, 41.20.Jb}

\section{Introduction}

The management of near-field radiative heat transfer (NFRHT) has promising
applications in energy conversion~\cite{Fio:18,Zha:17,Bas:09} and
information processing~\cite{Abd:17,Abd:16a}. Thanks to the
tunneling of evanescent waves (e.g., surface plasmon polaritions (SPPs)~\cite%
{Jou:05,Bor:15}, surface phonon polaritions (SPhPs)~\cite{She:09,Son:15}),
the radiative heat flux can exceed the blackbody limit by several orders
of magnitude for nanoscale separation~\cite{Vol:07,liu:15}. The field of
NFRHT grows rapidly in recent years~\cite{Cue:18}. Particularly, the thermal analogy of the electronic devices~\cite{Abd:17}, i.e.,thermotronics, provides a new possibility for information processing. The
fast thermal photon generated from NFRHT is an attractive information carrier
compared with its counterpart of the slow phonon (heat conduction)~\cite%
{Li:12}. Up to now, the proposals for fundamental circuit elements such as thermal
logic gates~\cite{Abd:16a}, thermal rectifiers~\cite{Ote:10}, thermal memory~%
\cite{Kub:14,Ito:16}, thermal transistors~\cite{Abd:14} have been reported.
Despite great progress in thermal circuit devices, the active and fast control of NFRHT is still highly desired for thermal information processing.

Recently, several strategies for active control of NFRHT have been
proposed~\cite{Ili:18,Hua:14,Cui:13,Kou:18,Abd:16b,Zhu:16,Eke:18,Mon:15,Van:11,Van:12,Gha:18}.
For instance, one can tune the NFRHT by external electric gating in 2D
materials~\cite{Ili:18} and ferroelectric materials~\cite{Hua:14}. Or one
can use optical pump, i.e., photo-excitation, to control NFRHT in chiral
meta-materials~\cite{Cui:13} and semiconductors~\cite{Kou:18}. In addition, one
can actively control NFRHT through an external magnetic field~\cite%
{Abd:16b,Zhu:16,Eke:18,Mon:15}. Examples such as near-field thermal Hall
effect~\cite{Abd:16b}, persistent directional heat current~\cite{Zhu:16},
anisotropic thermal magnetoresistance ~\cite{Eke:18}, and thermal modulation~%
\cite{Mon:15} were reported. Other schemes such as temperature control in
phase-change materials~\cite{Van:11,Van:12}, and mechanical strain~\cite%
{Gha:18} are also investigated. However, previous studies of various
strategies for active control of NFRHT have been focused on enhancing or
suppressing the total heat flux. Little attention was paid on the control of
multi-band spectral bands. Inspired by the multi-band devices in
telecommunications, a controllable thermal transfer system with multi-band
spectra could be significant for the development of thermal information
processing.

\begin{figure}[tbp]
\centerline{\includegraphics[width=8.0cm]{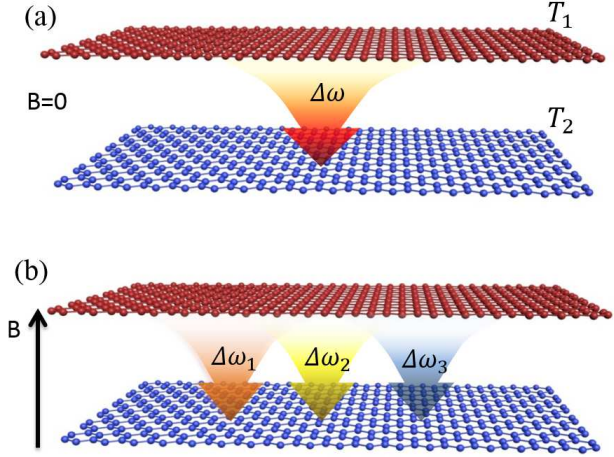}}
\caption{(color online) Schematic view of near-field radiative heat transfer
between two suspended graphene sheets in free space. The upper sheet with
higher temperature ($T_1$) radiates electromagnetic waves to the bottom one
with lower temperature ($T_2$). (a) The spectrum of radiative heat flux is
single-band in the absence of an external magnetic field. (b) The spectrum
becomes multi-band under a sufficient strong magnetic field.}
\end{figure}

In order to realize multi-band NFRHT, one straight scheme is to design a
system with multiple frequency-separated surface modes. The SPPs can be
modulated under an external magnetic field (or named magneto-plasmon)~\cite%
{Hu:15}. The multi-band dispersions, i.e., energy gaps
appear in the dispersion relationship, can be found in magneto-plasmon of
doped semiconductors~\cite{Bri:72} and graphene~\cite{Fer:12,Hu:14}.
Compared with the doped semiconductors, graphene is only one-atom thick~%
\cite{Gei:07}. Besides, the cyclotron energy of graphene, characterized by
the quantized Landau level (LL) transitions~\cite{Goe:11}, can be much larger
than those of semiconductors under the same magnetic field. The
magneto-plasmon of graphene lies in Terahertz and far infrared frequencies~\cite{Fer:12,Hu:14},
which guarantees a high-efficiency thermal excitation at room temperature.

In this work, we investigate the NFRHT between two parallel graphene sheets
under a perpendicular magnetic field. It is found that the single-band
spectrum for B=0 can be split into multi-band spectra under an external
magnetic field at room temperature. The multi-band spectra are determined by the LL transitions
of graphene, which show interesting properties due to thermal fluctuation.
For a modest magnetic field (e.g., B=4 T), dual-band spectra can be realized,
contributed from intra-band and inter-band LL transitions, respectively.
Interestingly, the latter one is quite robust against the change of chemical
potential. On the other hand, triple-band spectra can be found under a strong
magnetic field (e.g., B=15 T). Remarkably, there is an
additional band with the resonant peak appearing at near-zero frequency. The
great enhancement of NFRHT at such low frequency is attributed to the
magneto-plasmon zero modes. Finally, the properties of multi-band NFRHT due
to the scattering rate, separation distance and substrate effect are discussed at the end.

\section{Theories for magnetically tunable NFRHT}

The system considered in this work is depicted in Fig.~1. Two parallel graphene sheets
are suspended in free space with a separation distance $d$.
The temperatures for the upper and bottom sheets are respectively $T_1$ and $%
T_2$. The value of the perpendicular magnetic field B can be tuned though
external control. In this work, we consider the magnetic field $\text{B}<$17
T, where the Zeeman splitting is negligible~\cite{Zha:06}. The
magneto-optical conductivity of graphene is given by a tensor~\cite{Gus:07,
Fer:11}:

\begin{equation}
\left(
\begin{array}{cc}
\sigma _{xx} & \sigma _{xy} \\
\sigma _{yx} & \sigma _{yy}%
\end{array}%
\right) =\left(
\begin{array}{cc}
\sigma _{L} & \sigma _{H} \\
-\sigma _{H} & \sigma _{L}%
\end{array}%
\right),
\end{equation}%
where the subscripts $L$ and $H$ represent respectively the longitudinal and
Hall conductivities, and they can be written in compact forms ~\cite{Fer:11}%
:
\begin{equation}
\sigma _{L(H)}(\omega ,B)=g_{s}g_{v}\frac{e^{2}}{4h}\underset{n\neq m}{%
\overset{\infty }{\sum }}\frac{\Xi _{L(H)}^{nm}}{i\Delta _{nm}}\frac{%
n_{F}(E_{n})-n_{F}(E_{m})}{\hbar \omega +\Delta _{nm}+i\Gamma _{nm}},
\end{equation}%
where $g_{s}$=$g_{v}$=2 are respectively the spin and valley degeneracy
factors, $e$ is the charge of an electron, $h$ is the plank constant, $%
\Gamma _{nm}$ is the scattering rate (the broadening of LLs), $%
n_{F}(E_{n})=1/[1+\exp (E_n-\mu _{c})/k_{B}T]$ is the Fermi-Dirac
distribution, and $\mu _{c}$ is the chemical potential, $k_{B}$ is the
Boltzmann constant, $\Delta _{nm}=E_{n}-E_{m}$ is the LL energy transition
with $n$, $m$=0, $\pm 1$, $\pm 2$\ldots\ are the Landau indices, $E_{n}=$sign$(n)
[\hbar v_{F}/l_{B}]\sqrt{2|n|}$ is the energy of the $n$-th LL, $v_{F}$=$%
10^{6}$ m/s is the Fermi velocity, $l_{B}\equiv \sqrt{\hbar /(eB)}$ is the
magnetic length, and the auxiliary functions are given by:%
\begin{eqnarray}
\Xi _{L}^{nm} &=&\frac{\hbar ^{2}v_{F}^{2}}{l_{B}^{2}}(1+\delta
_{m,0}+\delta _{n,0})\delta _{|m|-|n|,\pm 1},  \notag \\
\Xi _{H}^{nm} &=&i\Xi _{L}^{nm}(\delta _{|m|,|n|-1}+\delta _{|m|-1,n}),
\end{eqnarray}
which are determined by the selection rule of LL transitions.
The radiative heat flux between two graphene sheets under an external
magnetic field can be calculated based on the fluctuation-dissipation theory
~\cite{Bie:11}:
\begin{equation}
\left\langle S\right\rangle =\frac{1}{(2\pi )^{2}}\int_{0}^{\infty
}\int_{0}^{\infty }\xi (\omega ,q)q\text{d}q[\Theta (\omega ,T_{1})-\Theta
(\omega ,T_{2})]\text{d}\omega ,
\end{equation}%
where $\Theta (\omega ,T)=\hbar \omega /(\exp (\hbar \omega /k_{B}T)-1)$ is
the average energy of a Planck's oscillator for angular frequency $\omega $
at temperature $T$, and $q$ is the wavevector parallel to the
surface planes, $\xi (\omega ,q)$ is
the energy transmission coefficient, which is expressed as follows:
\begin{eqnarray}
&&\xi (\omega ,q)\nonumber \\
&=&\left\{
\begin{array}{c}
\text{Tr}[(\mathbf{I}-\mathbf{R}_{2}^{\dagger }\mathbf{R}_{2})\mathbf{D}%
^{12}(\mathbf{I}-\mathbf{R}_{1}^{\dagger }\mathbf{R}_{1})\mathbf{D}%
^{12\dagger }], \\
\text{Tr}[(\mathbf{R}_{2}^{\dagger }-\mathbf{R}_{2})\mathbf{D}^{12}(\mathbf{R%
}_{1}^{\dagger }-\mathbf{R}_{1})\mathbf{D}^{12\dagger }]e^{-2|\gamma |d},%
\end{array}%
\right.
\begin{array}{c}
q<k_0 \nonumber \\
q>k_0 \nonumber
\end{array}\\
\end{eqnarray}%
where $\gamma =\sqrt{k_{0}^{2}-q^{2}}$ is the vertical wavevector with $%
k_{0}=\omega /c$ being the wavevector in vacuum, $\mathbf{D}^{12}=(\mathbf{I}%
-\mathbf{R}_{1}\mathbf{R}_{2}e^{2i\gamma d})^{-1}$ and $\mathbf{R}_{j}$ ($j$%
=1, 2) is the $2\times 2$ reflection matrix for the $j$-th graphene sheet,
having the form:
\begin{equation}
\mathbf{R}_{j}=\left(
\begin{array}{cc}
r_{j}^{ss} & r_{j}^{sp} \\
r_{j}^{ps} & r_{j}^{pp}%
\end{array}%
\right) ,
\end{equation}%
where the superscripts $s$ and $p$ represent respectively the polarizations
of transverse electric ($\mathbf{TE}$) and transverse magnetic ($\mathbf{TM}$%
) modes. The matrix element $r^{\alpha \beta }$ ($\alpha ,\beta $ = $s,p$)
represents the reflection coefficient for an incoming $\alpha $-polarized
plane wave being reflected into an outgoing $\beta $-polarized wave. For a
suspended graphene sheet under a perpendicular magnetic field, the
reflection coefficients are given analytically as follows~\cite%
{Kot:17,Tse:12}:

\begin{eqnarray}
r^{ss} &=&-\frac{2k_{0}^{2}\overline{\sigma }_{L}+k_{0}\gamma (\overline{%
\sigma }_{L}^{2}+\overline{\sigma }_{H}^{2})}{4k_{0}\gamma +(2\gamma
^{2}+2k_{0}^{2})\overline{\sigma }_{L}+k_{0}\gamma (\overline{\sigma }%
_{L}^{2}+\overline{\sigma }_{H}^{2})},  \notag \\
r^{pp} &=&-\frac{2\gamma ^{2}\overline{\sigma }_{L}+k_{0}\gamma (\overline{%
\sigma }_{L}^{2}+\overline{\sigma }_{H}^{2})}{4k_{0}\gamma +(2\gamma
^{2}+2k_{0}^{2})\overline{\sigma }_{L}+k_{0}\gamma (\overline{\sigma }%
_{L}^{2}+\overline{\sigma }_{H}^{2})}, \\
r^{ps} &=&-r^{sp}=\frac{2k_{0}\gamma \overline{\sigma }_{H}}{4k_{0}\gamma
+(2\gamma ^{2}+2k_{0}^{2})\overline{\sigma }_{L}+k_{0}\gamma (\overline{%
\sigma }_{L}^{2}+\overline{\sigma }_{H}^{2})},  \notag
\end{eqnarray}
where the elements of conductivity tensor are normalized by the free-space
impedance $\sqrt{\mu _{0}/\varepsilon _{0}}$. Note that the denominator in all
reflection coefficients is the same. By setting this denominator to be zero and the loss term $\Gamma $=0, the dispersion of magneto-plasmon in single graphene sheet can be obtained.

We now consider the NFRHT in the low frequency limit $\omega \rightarrow 0$, where the normalized conductivities Im$\overline{\sigma }_{L}$, Im$\overline{%
\sigma }_{H}\rightarrow 0$ and Re$\overline{\sigma }_{L}$, Re$\overline{%
\sigma }_{H}\ll 1$. For very large wave-vector $q\gg k_{0}$, i.e., $|\gamma
|\gg k_{0}$, only the reflection coefficient $r^{pp}$ plays an important
role, and it can be simplified as:
\begin{equation}
r^{pp}=-\frac{iq\overline{\sigma }_{L}}{2k_{0}+iq\overline{\sigma }_{L}}.
\end{equation}
The pole of the denominator (real part) appears at $\omega $=0, which
corresponds to a magneto-plasmon zero-mode. The NFRHT is proportional to Im$%
[r^{pp}]$ ~\cite{Ili:18}, where it has the expression Im[$%
r^{pp}$]=-2$k_{0}/(q\mathrm{Re}\overline{\sigma }_{L})$. For a strong magnetic field, the longitudinal conductivity of graphene, i.e., the term $\mathrm{Re}\overline{\sigma }_{L}$, can be diminished with orders of magnitude compared with the non-magnetic configuration, based on the calculations of Eq. (2). As a result, the NFRHT at near-zero frequency can be enhanced greatly due to the magneto-plasmon zero modes, which is an important finding in our work. It should be pointed out that the magneto-plasmon zero modes in 2D electron gas
systems were also demonstrated in the Ref.~\cite{Jin:16}.

\section{Results and discussions}

\begin{figure}[tbp]
\centerline{\includegraphics[width=8.0cm]{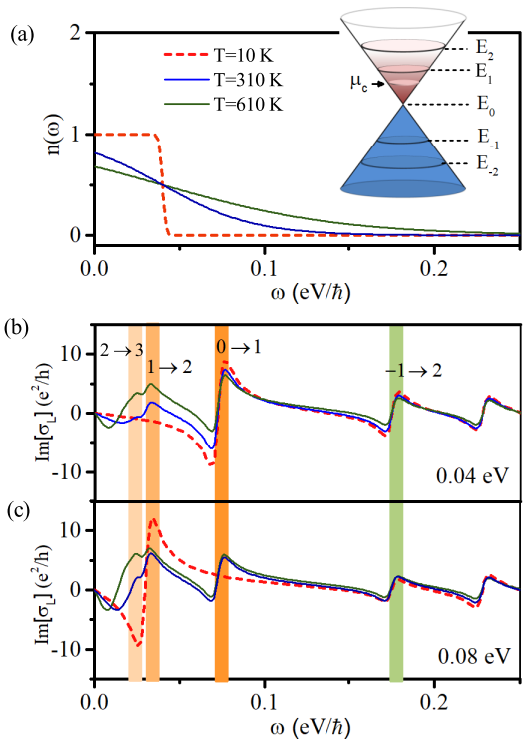}}
\caption{(color online) (a) The Fermi-Dirac distribution of graphene for
different temperatures. The inset illustrates the band structure of a
graphene sheet with chemical potential $\protect\mu _{c}$ below the first LL. The
imaginary part of longitudinal conductivity Im[$\protect\sigma _{L}$] for
(b) $\protect\mu _{c}$=0.04 eV and (c) $\protect\mu _{c}$=0.08 eV. The
shaded regimes in (b) and (c) indicate the peaks of conductivity contributed
by different LL transitions.  The scattering rate is set to be a constant with $\Gamma _{nm}$=$\Gamma $= 4 meV. The external magnetic field B=4 T, and the first LL  E$_{1}\sim $0.07 eV.
}
\end{figure}

\begin{figure*}[tbp]
\centerline{\includegraphics[width=14cm]{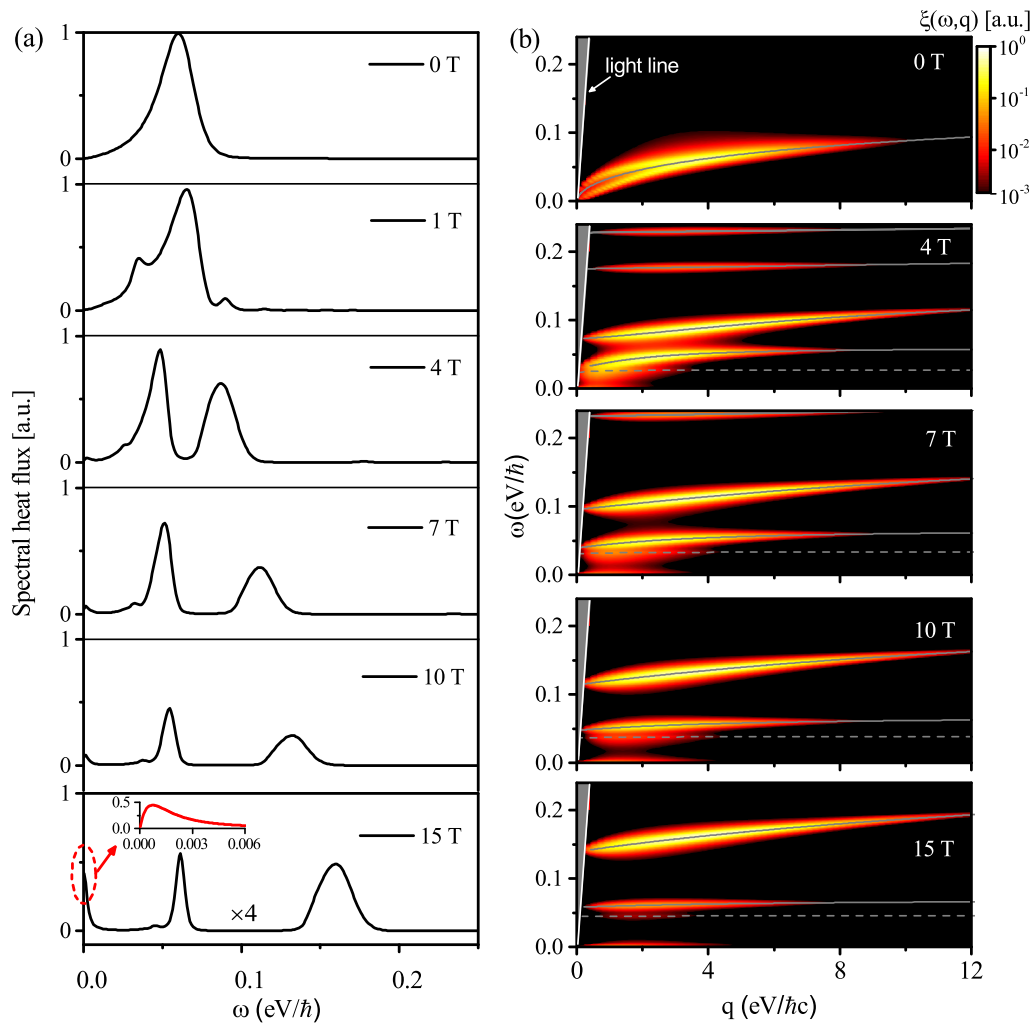}}
\caption{(color online) (a) Spectral heat flux and (b)
transmission coefficients(log scale) for two parallel graphene sheets under different
values of magnetic field. The inset in (a) shows the spectrum near-zero
frequency for B=15 T. In (b), the white line illustrates the light line q=$\omega$/c, and the gray lines represent the corresponding dispersions of magneto-plasmon for single graphene sheet. The dashed
gray lines stand for the dispersions generated from the intra-band transition E$_{2}\rightarrow $E$_{3}$. We set the
chemical potential for two graphene sheets $\protect\mu_{c1}=\protect\mu%
_{c2} $=0.08 eV, the scattering rate $\Gamma$=4 meV, the high temperature $%
T_1$=310 K, the low temperature $T_2$=$T_1$-10 K, and the separation
distance $d$=100 nm. }
\label{Fig:fig3}
\end{figure*}
To understand the magneto-plasmon of graphene at high temperature, we firstly inspect the Fermi-Dirac distribution $n(\omega )$ in Fig.~2(a). Clearly, $n(\omega )$ is a good quantum number at low temperature 10 K, whereas it becomes partially occupied
at high temperatures 310 K and 610 K, due to the thermal fluctuation. The
inset schematically shows the band structure of graphene with quantized LLs
and Fermi-Dirac distribution of electron at high temperatures. The
long-decay tails of $n(\omega )$ enable multiple intra-band LL transitions
take place (such as E$_{1}\rightarrow $E$_{2}$, E$_{2}\rightarrow $E$_{3}$)
even though $\mu _{c}$ is smaller than the first LL E$_{1}$.

We are interest in the imaginary part of longitudinal conductivity, i.e., Im[%
$\sigma _{L}$], as the magneto-plasmon of graphene is mainly determined by
this component \cite{Fer:12}. The magneto-plasmon of graphene are confirmed strongly when Im[$\sigma _{L}$] is a positive number, providing a capability for the enhancement of NFRHT.
The comb-like curves of Im[$\sigma _{L}$] are shown in Figs.~2(b) and 2(c), where the magnetic field B =4 T, and
the first LL  E$_{1}\sim $0.07 eV. The values of chemical potential of graphene
are controllable though external electric gating~\cite{Gri:12}.  For chemical potential $\mu _{c}$=0.04 eV$<$ E%
$_{1}$,  there is no intra-band transition for temperature 10 K%
. However,  multiple intra-band transitions (mainly E$_{1}\rightarrow $E$_{2}$ and E$%
_{2}\rightarrow $E$_{3}$) take place for 310 K and 610 K at low frequency, which differ
considerably from those of 10 K.  The conductivity Im[$\sigma _{L}$] with chemical potential located between E$_{1}$ and E$_{2}$ is shown in Fig.~2(c) for $\mu _{c}$=0.08 eV. At this time, the conductivity at low temperature 10 K is
somewhat different from those of $\mu _{c}$=0.04 eV. Now there is an
intra-band transition E$_{1}\rightarrow $E$_{2}$, whereas the inter-band
transition E$_{0}\rightarrow $E$_{1}$ is absent. By contrast, both multiple
intra-band (E$_{1}$$\rightarrow$ E$_{2}$, E$_{2}$ $\rightarrow$ E$_{3}$) and
inter-band (E$_{0}$ $\rightarrow$ E$_{1}$) LL transitions still exist at temperatures 310 K and 610 K.
However, the dependence of Im[$\sigma _{L}$] on the temperature becomes small at high frequency. This is because the energies of inter-band LLs transitions are much higher
than those of thermal fluctuation (e.g, $k_{B}T\sim $0.027 eV for 310 K).

\begin{figure*}[tbp]
\centerline{\includegraphics[width=14cm]{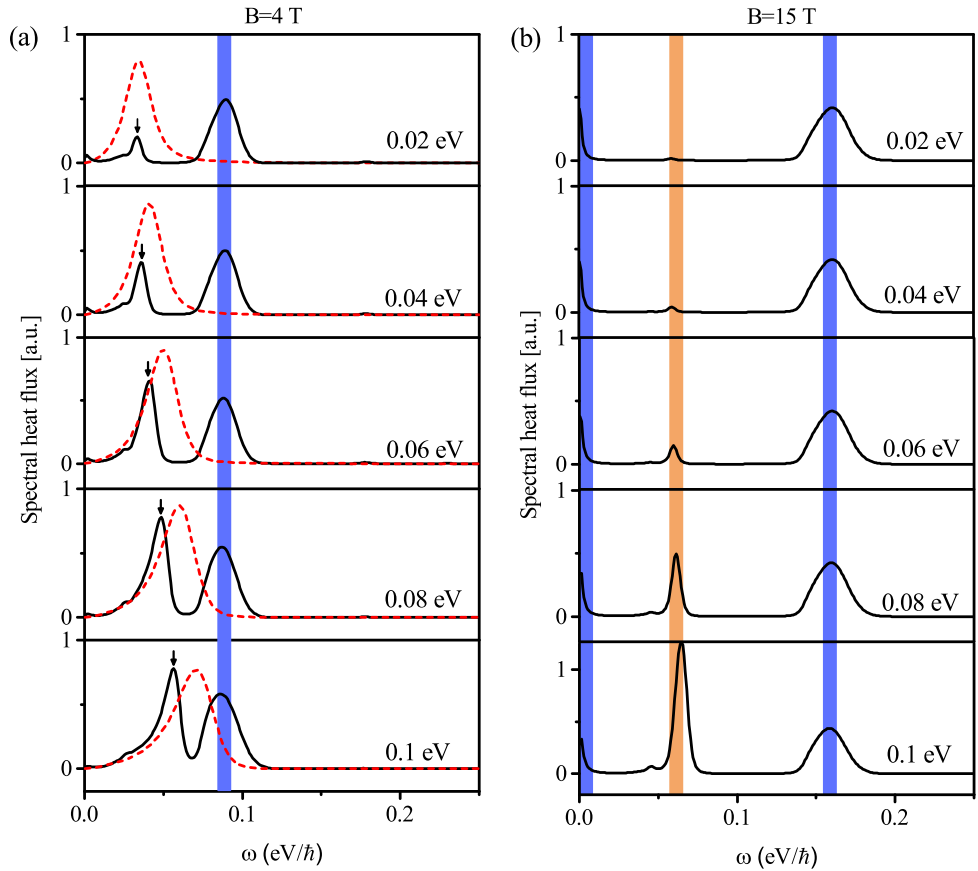}}
\caption{(color online) Spectral heat flux via chemical
potential ($\protect\mu_{c1}$=$\protect\mu_{c2}$) for (a) B=4 T. The red
dashed curves represent the corresponding spectra for B=0, where the resonant
peaks have a blue shift as the chemical potential increases. (b) B=15 T.
The middle spectral band (the orange shaded regime) changes dramatically with increasing
the chemical potential. However, the spectral bands marked by the
blue shaded regimes in (a) and (b) are almost unchanged, showing a robust feature. }
\label{Fig:fig4}
\end{figure*}

Figure 3(a) shows the spectra of radiative heat flux under different magnetic fields. In the absence of magnetic field, the spectral heat flux is single-band as reported in Ref.~\cite{Ili:12}. Compared with the non-magnetic configuration, the spectra are modulated correspondingly after the magnetic field is turned on. For instance, there are some small peaks for B=1 T. As the magnetic field increases further, dual-band spectra can be found for
B=4 T, 7 T and 10 T. Overall, the peaks of the two spectral bands undergo a blue shift and
the energy gap between these two peaks enlarges as the magnetic field increases. For a strong
magnetic field B=15 T, triple-band spectrum can be seen clearly. In addition to those two spectral bands mentioned above, there is a remarkable band close to zero frequency, and it is plotted in the inset for clarity. The resonant peak for this low-frequency band
appears at 0.8 meV ($\sim$200 GHz), which lies in the microwave
regime. The great enhancement of NFRHT in such low frequency has not been
revealed in any previous systems before. Although the amplitude of spectrum
has been multiplied 4 times for B=15 T, the radiative heat flux
can still exceed the black body limit over 40 times, which is large enough
for potential applications in triple-band thermal devices.

To clarify the origin of multi-band spectra, we plot of transmission
coefficient $\xi (\omega ,q)$ below the light line in Fig.~3(b). The large wave-vector $q\gg k_0$ indicates that the NFRHT is mainly contributed by the evanescent waves. Specifically, the transmission coefficient for B=0 is continuous over the frequency regime, resulting in a single-band spectrum.
By contrast, the transmission coefficients for B=4 T, 7 T, 10 T and 15 T
turn out to be frequency separated configurations, leading to multi-band
spectra. Interestingly, there is an extraordinary band started from zero
frequency, which is generated from magneto-plasmon zero modes as mentioned
before. This band can be prominently seen, especially for strong magnetic
field B=10 T and 15 T. It is worth mentioning that the transmission coefficient $\xi (\omega ,q)$ due to high-order
LL transitions can be seen as well for B=4 T and 7 T with $\omega>$0.15 eV/$\hbar$. However, the contribution of high-order
LL transitions to the NFRHT is very small, due to the fast decay of the term
$\Theta (\omega ,T)$ at high frequency. For comparison, the dispersions of magneto-plasmon in single graphene sheet are given correspondingly. It is found that the dispersions turn out to be a series of branches at the presence of magnetic field, which agree well with transmission coefficient $\xi (\omega ,q)$. Since the transmission coefficient $\xi
(\omega ,q)$ generated from intra-band transition E$_{2}\rightarrow $E$%
_{3}$ is very small, the corresponding dispersions are plotted in the dashed
gray lines to indicate their insignificance in NFRHT.

The chemical potential of graphene is an important parameter, which can be tuned greatly by electrical gating. The spectral heat flux via different chemical potential are shown in Figs.~4(a) and~4(b). For B= 4 T, there
are two primary spectral bands. The first primary band (marked by the black arrows) has a relative poor excitation when the chemical potential is small (e.g., $\mu_c=0.02$ eV). As the chemical potential increases, however, the magnitudes of this band increase and the spectral peaks have a blue shift. This evolution is attributed to the properties of intra-band transition. On the other hand, the second band (marked by blue shaded regime) is contributed by the inter-band LL transition (E$_0$$\rightarrow $E$_1$). Surprisingly, the
magnitude and line-shape of this spectral band have little change as
chemical potential increases from 0.02 eV to 0.1 eV, showing a robust
feature.  For comparison, the spectral heat flux for B=0 are given
by the red dashed lines. The resonant peaks of spectra for B=0 undergo a blue
shift when the chemical potential increases, indicating a trivial feature
contributed by the intra-band transitions. For a strong magnetic field B=15
T, the first and the third spectral bands (blue shaded regimes) are robust
against the change of chemical potential as well.
Interestingly, the middle spectral band (orange shaded regimes) contributed
by intra-band transitions is suppressed strongly at low doping 0.02
eV and 0.04 eV. This is attributed to high energy of first LL (E$_{1}$$\sim $ 0.14 eV
for 15 T), resulting in low probability of intra-band transition. As
expected, the magnitude of the middle spectrum-band increases with the
chemical potential, and it has the same order with the two side
bands when the chemical potential $\mu _{c}\sim$0.08 eV. Thus, triple-band
radiative heat transfer can be realized for suitable tuning of chemical
potential under a strong magnetic field. As the chemical potential increases
further to $\mu _{c}$=0.1 eV, the middle band is dominant.

\begin{figure}[tbp]
\centerline{\includegraphics[width=8.2cm]{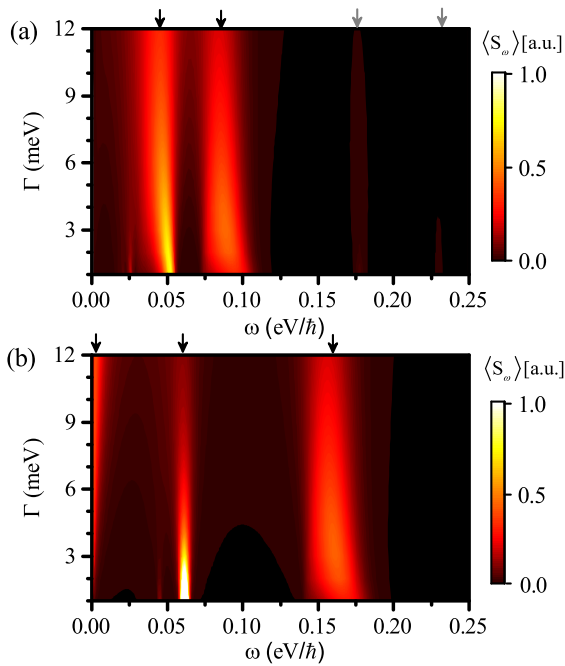}}
\caption{(color online) Contour plots of spectral heat flux for different
scattering rate. (a) B=4 T, and (b) B=15 T. The black arrows indicates the
primary bands, and the gray arrows represent secondary bands contributed
from high order LL transitions. The chemical potential, temperatures and
separation distance are kept the same as those in Fig. 3. }
\label{Fig:fig5}
\end{figure}

\begin{figure}[tbp]
\centerline{\includegraphics[width=8.3cm]{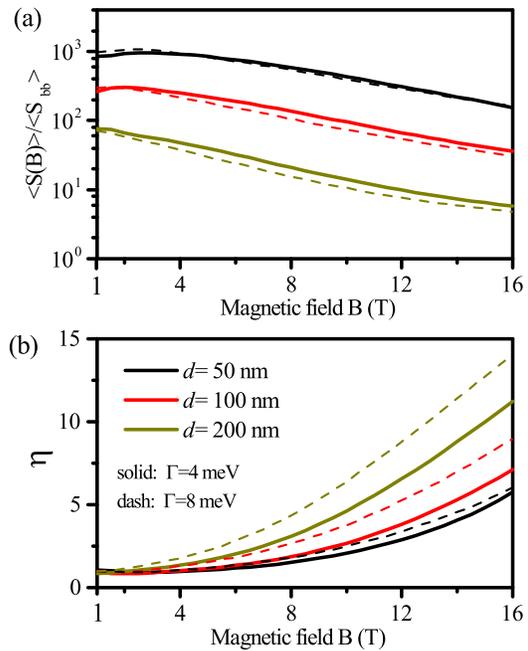}}
\caption{(color online) (a) The radiative heat flux as a function of
magnetic field, normalized by blackbody limit. (b)The modulation of
radiative heat flux in the presence of magnetic field, defined
as $\eta=\left\langle S(\text{B}=0)\right\rangle $/$\left\langle S(\text{B}%
)\right\rangle $. The solid and dash curves represent the scattering rate $%
\Gamma$=4 meV and 8 meV, respectively. The chemical potential and
temperatures are kept the same as those in Fig. 3.}
\label{Fig:fig6}
\end{figure}

To study in a more general way, the scattering rate is assumed to be a constant but within a typical range $\Gamma\in$ [1, 12]meV (see, e.g., \cite{Fer:12,Cra:11}). The contour plots of spectral heat flux are shown in Figs.~5(a) and~5(b).
For B=4 T, the peak frequencies of the two primary bands indicated by the black arrows almost
do not changed as $\Gamma $ increase from 1 meV to 12 meV.
Although the magnitude of spectral band become smaller for large scattering
rate, the two primary bands are still dominant.
The contour plot of spectral heat flux for B=15 T is shown in Fig.~5(b).
Again, three primary bands can be seen clearly even for large scattering
rate. The magnitude of the middle band decreases monotonously as the
scattering rate increases. Interestingly, the magnitude for the
low-frequency band is almost unchanged, whereas the bandwidths become
broaden with increasing scattering rate.

Figure 6(a) shows the radiative heat flux as a function of magnetic field for scattering rate $\Gamma$=4 meV and 8 meV. It is found that the deviation of $\left\langle S(\text{B})\right\rangle $ for these two scattering rate is small for separation distance $d$=50 nm. For $d$=50 nm and $d$=100 nm, there exist a maximum enhancement of spectral heat flux near B$\sim$2
T, and the values are about 1000 times and 300 times over the blackbody limit, respectively. Nevertheless, $%
\left\langle S(\text{B})\right\rangle $ decreases monotonically as the value
of magnetic field increases further. For $d$=200 nm, however, $\left\langle
S(B)\right\rangle $ decreases monotonically from B=1 T to B=16 T. The
modulation of NFRHT due to the presence of external magnetic field is shown
in Fig. 6(b), where the ratio $\eta=\left\langle S(\text{B}=0)\right\rangle $/$%
\left\langle S(\text{B})\right\rangle $ is used. This ratio is relatively
small for a moderate magnetic field, whereas it can increase dramatically as
the magnetic field becomes strong. Taking B=15 T ($\Gamma $=4 meV) as an
example, the ratio $\eta$ is about 4.8, 6.2 and 10 times for separation
distances $d$=50 nm, 100 nm and 200 nm, respectively. The ratio $\eta$ can be
further enlarged as the scattering rate $\Gamma$=8 meV. The modulation of
NFRHT between graphene sheets here is much larger than those of semiconductor
such as InSb and doping Si under a perpendicular magnetic field~\cite{Mon:15}%
.

\begin{figure}[tbp]
\centerline{\includegraphics[width=7.5cm]{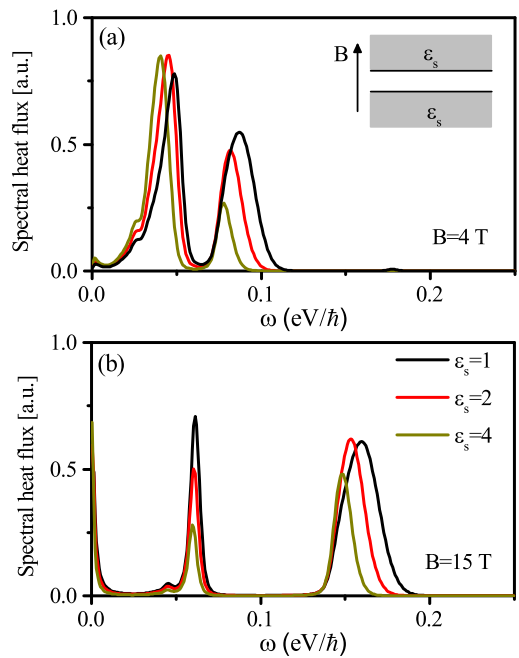}}
\caption{(color online) Spectral heat flux for graphene sheets placed on top of substrates with permittivity
$\varepsilon_s$, where the thickness of the substrates is semi-infinite. The
magnetic fields are (a) 4 T and (b) 15 T, respectively. Other parameters are kept the same
as those in Fig.~3.}
\label{Fig:fig7}
\end{figure}

Technically, suspended graphene sheets can be realized in the experiments~%
\cite{Mey:07,Bol:08}. However, graphene sheets are generally deposited on a
variety of different substrates~\cite{Edw:13}. The substrate effect for NFRHT is demonstrated in Figs.~7(a)
and 7(b). As the permittivity of the dielectric substrate (non-polar and non-dispersive
materials) $\varepsilon_s$ increases, all spectral peaks undergo a red shift. Besides, the amplitudes of the second spectral band for B=4 T and the middle spectral band for B=15 T decrease accordingly. It is expected that the substrate effect can be diminished as the thickness of substrates is reduced. Nevertheless, the multi-band spectra still persist
even the substrates are taken into account. Finally, it is worth mentioning that the type of doping (electron or hole doping) in graphene could determine the sign of Hall conductivity $\sigma _{H}$. The sign of $\sigma _{H}$ is important for some
remarkable phenomena such as quantum Casimir effect~\cite{Tse:12} and Faraday
rotation~\cite{Cra:11}. However, the sign of $\sigma _{H}$ has little effect
for NFRHT since the magneto-plasmon of graphene is dominant by the
longitudinal component of conductivities.

\section{Conclusions}

In summary, magnetically tunable multi-band NFRHT is revealed in a system of
two suspended graphene sheets. The single-band spectra for B=0 can be modulated greatly in the presence of a sufficient strong magnetic-field. Dual-band spectra can be realized for moderate magnetic fields (e.g., B=4 T). One band is determined by intra-band transitions in the classical regime, which undergoes a blue shift as the chemical potential
increases. The other band is contributed by inter-band LL transition E$%
_0\rightarrow$E$_1$ in the quantum regime, showing a robust feature against
the change of chemical potential. On the other hand, triple-band spectra can
be found under a strong magnetic field (e.g., B=15 T). Remarkably, there is
an extremely low-frequency band with the resonant peak appearing at microwave regime, stemming from the magneto-plasmon zero modes. The performances of NFRHT due
to the scattering rate, separation distance and substrate effect are demonstrated as well. Although the magnetic field considered in our work is
static, it can indeed be time-dependent such as sin-cos, sawtooth,
square-wave etc. As a result, a dynamic tunable multi-band NFRHT would be
possible under a time-dependent magnetic field, providing a possibility for multi-band information
transfer.

\begin{acknowledgments}
This work is supported by the National Natural Science Foundation of China
(Grant No. 11747100, 11704254, 11804288), and the Innovation Scientists and
Technicians Troop Construction Projects of Henan Province. The research of
L. X. Ge is further supported by Nanhu Scholars Program for Young Scholars
of XYNU. Work in Saudi Arabia was supported by King Abdullah University of Science and Technology (KAUST) Baseline Research Fund BAS/1626-01-01.
\end{acknowledgments}


\begin{thebibliography}{99}
\bibitem{Fio:18} A. Fiorino, L. Zhu, D. Thompson, R. Mittapally, P. Reddy
and E. Meyhofer,  Nat.
Nanotechnol. $\mathbf{13}$, 806-811 (2018).

\bibitem{Zha:17} B. Zhao, K. Chen, S. Buddhiraju, G. Bhatt, M. Lipson, S.
Fan,  Nano Energy $\mathbf{41}$, 344-350 (2017).

\bibitem{Bas:09} S. Basu, Z. M. Zhang, and C. J. Fu,  Int. J. Energy
Res. $\mathbf{33}$, 1203-1232 (2009).

\bibitem{Abd:17} P. B.-Abdallah and S.-A. Biehs,  Z. Naturforsch. A $\mathbf{72}$%
(2), 151-162 (2017).

\bibitem{Abd:16a} P. B.-Abdallah and S.-A. Biehs, Phys. Rev. B $\mathbf{94}$, 241401(R)
(2016).

\bibitem{Jou:05} K. Joulain, J.-P. Mulet, F. Marquier, R. Carminati, and
J.-J. Greffet,  Surf. Sci. Rep. $\mathbf{57}$, 59 (2005).

\bibitem{Bor:15} S. V. Boriskina, J. K. Tong, Y. Huang, J. Zhou, V.
Chiloyan, G. Chen, Photonics $\mathbf{2}$, 659-683 (2015).

\bibitem{She:09} S. Shen, A. Narayanaswamy and G. Chen,  Nano Lett. $%
\mathbf{9}$ (8), 2909-2913(2009).

\bibitem{Son:15} B. Song, Y. Ganjeh, S. Sadat, D. Thompson, A. Fiorino, V.
Fern\'{a}ndez-Hurtado, J. Feist, F. J. Garcia-Vidal, J. C. Cuevas, P. Reddy
and E. Meyhofer,  Nat. Nanotechnol. $\mathbf{10}$, 253-258 (2015).

\bibitem{Vol:07} A. I. Volokitin and B. N. J. Persson,  Rev. Mod. Phys. $\mathbf{%
79}$, 1291 (2007).

\bibitem{liu:15} X. Liu, L. Wang, and Z. M. Zhang,  Nanosc. Microsc. Therm. $\mathbf{%
19}$, 98-126 (2015).

\bibitem{Cue:18} J. C. Cuevas, and Francisco J. Garcia-Vidal,  ACS Photonics   $\mathbf{5}$(10), 3896-3915 (2018).

\bibitem{Li:12} N. Li, J. Ren, L. Wang, G. Zhang, P. H$\ddot{a}$nggi, B. Li ,
 Rev. Mod. Phys. $\mathbf{84}$ (3), 1045 (2012).

\bibitem{Ote:10} C. R. Otey, W. T. Lau, and S. Fan,  Phys. Rev. Lett. $\mathbf{104}$, 154301(2010).

\bibitem{Kub:14} V. Kubytskyi, S.-A. Biehs, and P. B.-Abdallah,  Phys. Rev. Lett. $\mathbf{113}$, 074301
(2014).

\bibitem{Ito:16} K. Ito, K. Nishikawa, and H. Iizuka,  Appl. Phys. Lett. $\mathbf{108}$, 053507 (2016).

\bibitem{Abd:14} P. B.-Abdallah and S.-A. Biehs,  Phys. Rev. Lett. $\mathbf{112}$, 044301(2014).

\bibitem{Ili:18} O. Ilic, N. H. Thomas, T. Christensen, M. C. Sherrott, M.
Solja$\breve{c}$i$\acute{c}$, A. J. Minnich, O. D. Miller, and H. A. Atwater,  ACS Nano $\mathbf{12}$
(3), 2474-2481(2018).

\bibitem{Hua:14} Y. Huang, S. V. Boriskina, G. Chen.  Appl. Phys. Lett.$\mathbf{105}$, 244102 (2014).

\bibitem{Cui:13} L. Cui, Y. Huang, J. Wang, and K.-Y. Zhu,  Appl. Phys. Lett. $\mathbf{102}$, 053106 (2013).

\bibitem{Kou:18} J. Kou and A. J. Minnich,  Opt. Express $\mathbf{26}$, A729-A736
(2018).

\bibitem{Abd:16b} P. B.-Abdallah,  Phys. Rev. Lett. $\mathbf{116}$, 084301 (2016).

\bibitem{Zhu:16} L. Zhu, and S. Fan,  Phys. Rev. Lett. $\mathbf{117}$, 134303 (2016).

\bibitem{Eke:18} R. M. Ab. Ekeroth, P. B.-Abdallah, J. C. Cuevas, and A. Garc%
\'{\i}a-Mart\'{\i}n,  ACS Photonics $\mathbf{5}$, 705-710
(2018).

\bibitem{Mon:15} E. Moncada-Villa, V. Fern$\acute{a}$ndez-Hurtado, F. J. Garc\'{\i}a-Vidal, A. Garc\'{\i}a-Mart\'{\i}n, and J.
C. Cuevas,  Phys. Rev. B $\mathbf{92}$, 125418 (2015).

\bibitem{Van:11} P. J. van Zwol, K. Joulain, P. B.-Abdallah, J. J. Greffet,
and J. Chevrier,  Phys. Rev. B $\mathbf{83}$, 201404(R) (2011).

\bibitem{Van:12} P. J. van Zwol, L. Ranno, and J. Chevrier,  Phys. Rev. Lett. $\mathbf{108}$, 234301 (2012).

\bibitem{Gha:18} A. Ghanekar, M. Ricci, Y. Tian, O. Gregory, and Y. Zheng,
 Appl. Phys. Lett. $\mathbf{112}$, 241104 (2018).

\bibitem{Hu:15} B. Hu, Y. Zhang and Q. J. Wang,  Nanophotonics $\mathbf{%
4}$, 383-396 (2015).

\bibitem{Bri:72} J. J. Brion, R. F. Wallis, A. Hartstein and E. Burstein, Phys. Rev. Lett. $%
\mathbf{28}$, 1455 (1972).

\bibitem{Fer:12} A. Ferreira, N. M. R. Peres, and A. H. Castro Neto, Phys. Rev. B $\mathbf{85}$,
205426 (2012).

\bibitem{Hu:14} B. Hu, J. Tao, Y. Zhang, and Q. J. Wang, Opt. Express $%
\mathbf{22}$, 21727 (2014).

\bibitem{Gei:07} A. K. Geim and K. S. Novoselov, 
Nat. Mater. $\mathbf{6}$, 183 (2007).

\bibitem{Goe:11} M. O. Goerbig,  Rev. Mod. Phys. $\mathbf{83}$, 1193 (2011).

\bibitem{Zha:06} Y. Zhang, Z. Jiang, J. P. Small, M. S. Purewal, Y.-W. Tan, M. Fazlollahi, J. D. Chudow, J. A. Jaszczak, H. L. Stormer, and P. Kim,  Phys. Rev. Lett. $\mathbf{96}$, 136806 (2006).

\bibitem{Gus:07} V. P. Gusynin, S. G. Sharapov, and J. P. Carbotte,
J. Phys.: Condens. Matter $%
\mathbf{19}$, 026222 (2007).

\bibitem{Fer:11} A. Ferreira, J. Viana-Gomes, Yu. V. Bludov, V. Pereira, N.
M. R. Peres, and A. H. Castro Neto,  Phys. Rev. B $\mathbf{84}$, 235410
(2011).

\bibitem{Bie:11} S. A. Biehs, P. Ben-Abdallah, F. S. S. Rosa, K. Joulain,
and J.-J.Greffet,  Opt. Express $\mathbf{19}$, A1088 (2011).

\bibitem{Kot:17} O. V. Kotov and Yu. E. Lozovik,  Phys. Rev. B $\mathbf{96}$, 235403 (2017).

\bibitem{Tse:12} W.-K. Tse and A. H. MacDonald,
Phys. Rev. Lett. $\mathbf{109}$, 236806 (2012).

\bibitem{Jin:16} D. Jin, L. Lu, Z. Wang, C. Fang, J. D. Joannopoulos, M.
Solja$\breve{c}$i$\acute{c}$, L. Fu and N. X. Fang,
Nat. Commun. $\mathbf{7}$, 13486 (2016).

\bibitem{Gri:12} A. N. Grigorenko, M. Polini and K. S. Novoselov,  Nat. Photonics $\mathbf{6}$, 749-758 (2012).

\bibitem{Ili:12} O. Ilic, M. Jablan, J. D. Joannopoulos, I. Celanovic, H.
Buljan, and M. Solja$\breve{c}$i$\acute{c}$,  Phys. Rev. B $\mathbf{85}$, 155422 (2012).

\bibitem{Cra:11} I. Crassee, J. Levallois, A. L. Walter, M. Ostler, A.
Bostwick, E. Rotenberg, T. Seyller, D. van der Marel and A. B. Kuzmenko,
 Nat. Phys. $\mathbf{7}$, 48-51 (2011).

\bibitem{Mey:07} J. C. Meyer, A. K. Geim, M. I. Katsnelson, K. S. Novoselov,
T. J. Booth and S. Roth, Nature $\mathbf{446}$, 60-63 (2007).

\bibitem{Bol:08} K. I. Bolotin, K. J. Sikes, Z. Jiang, M. Klima, G.
Fudenberg, J. Hone, P. Kim and H. L. Stormer,  Solid State Commun. $\mathbf{146}$, 351-355 (2008).

\bibitem{Edw:13} R. S. Edwards and K. S. Coleman,  Nanoscale $\mathbf{5}$, 38 (2013).



\end{thebibliography}
\end{document}